\documentclass[preprint,12pt,3p]{elsarticle}

\usepackage{amssymb}
\usepackage{graphicx}
\usepackage{amsmath,amssymb}
\usepackage{stmaryrd} 
\usepackage{cases}
\usepackage{subcaption}
\usepackage{verbatim}

\graphicspath{{.}{fig/}{data/}}

\usepackage{moreverb,url}
\usepackage[colorlinks,bookmarksopen,bookmarksnumbered,citecolor=red,urlcolor=red]{hyperref}

\usepackage{tikz}
\usetikzlibrary{patterns}
\usetikzlibrary{positioning}
\usetikzlibrary{arrows,decorations.markings}

\usepackage[colorinlistoftodos]{todonotes}

\journal{Computers \& Operations research}

\begin{document}

\begin{frontmatter}

\title{Post-Prognostics Decision for Optimizing the Commitment of Fuel
  Cell Systems}

\author[label1]{St\'ephane Chr\'etien\corref{cor1}\fnref{label3}}
\address[label1]{National Physical Laboratory}

\cortext[cor1]{Corresponding author}

\ead{stephane.chretien@npl.co.uk}
\ead[url]{https://sites.google.com/view/stephanechretien/home}

\author[label5]{Nathalie Herr}
\address[label5]{FEMTO-ST/AS2M, Univ. Bourgogne Franche-Comt\'e/UFC/CNRS/ENSMM,Besan ̧con, France}
\ead{nathalie.herr@femto-st.fr}

\author[label5]{Jean-Marc Nicod}
\ead{jean-marc.nicod@femto-st.fr}

\author[label5]{Christophe Varnier}
\ead{christophe.varnier@femto-st.fr}

\begin{abstract}
  In a post-prognostics decision context, this paper addresses the problem of maximizing the useful life of a platform composed of several parallel machines under service constraint. Application on multi-stack fuel cell systems is considered. In order to propose a solution to the insufficient durability of fuel cells, the purpose is to define a commitment strategy by determining at each time the contribution of each fuel cell stack to the global output so as to satisfy the demand as long as possible. A relaxed version of the problem is introduced, which makes it potentially solvable for very large instances. Results based on computational experiments illustrate the efficiency of the new approach, based on the Mirror Prox algorithm, when compared with a simple method of successive projections onto the constraint sets associated with the problem.
\end{abstract}


\end{frontmatter}

\newcommand\BibTeX{{\rmfamily B\kern-.05em \textsc{i\kern-.025em b}\kern-.08em
T\kern-.1667em\lower.7ex\hbox{E}\kern-.125emX}}

\newcommand{\toreview}[1]{\todo[color=red!50!blue,inline]{\textcolor{white}{OK
      pour review - #1}}}

\newcommand{\PHM}{\ensuremath{\text{\textit{PHM}}}}
\newcommand{\RUL}{\ensuremath{\mbox{\textit{RUL}}}}
\newcommand{\EOL}{\ensuremath{\mbox{\textit{EOL}}}}
\newcommand{\UB}{\ensuremath{\text{\textit{\textsc{UB}}}}}

\let\oldexists\exists
\renewcommand{\exists}{\, \oldexists \, }
\let\oldforall\forall
\renewcommand{\forall}{\, \oldforall\, }
\newcommand{\tq}{\ \ensuremath{\mbox{s.t.}} \,}
\newcommand{\argmin}{\mathop{\mathrm{argmin}}}

\newcommand{\Pmin}{P\ensuremath{\text{min}}}
\newcommand{\Pmax}{P\ensuremath{\text{max}}}
\newcommand{\Pnom}{P\ensuremath{\text{nom}}}
\newcommand{\Popt}{P\ensuremath{\text{opt}}}
\newcommand{\Ptot}{P\ensuremath{\text{tot}}}
\newcommand{\fmax}{\ensuremath{\textit{f}\text{max}}}
\newcommand{\Fmax}{\ensuremath{\textit{F}\text{max}}}
\newcommand{\fmin}{\ensuremath{\textit{f}\text{min}}}
\newcommand{\RULopt}{\ensuremath{\textit{RUL}\text{opt}}}
\newcommand{\RULmin}{\ensuremath{\textit{RUL}\text{min}}}
\newcommand{\RULmax}{\ensuremath{\textit{RUL}\text{max}}}
\newcommand{\tstart}{\ensuremath{\textit{t}\text{start}}}
\newcommand{\tend}{\ensuremath{\textit{t}\text{end}}}
\newcommand{\distfmax}{\ensuremath{\textit{distf}\text{max}}}
\newcommand{\inc}{\ensuremath{\textit{inc}}}

\newcommand{\blockParam}[7]{\node[shape=rectangle, text=#4, text centered,
  draw, rounded corners = 1mm, fill=#3, text width=#5, minimum
  height=1.2cm, rotate=#7] (#1) at #6 {#2};}

\definecolor{OliveGreen}{cmyk}{0.64,0,0.95,0.40} 
\definecolor{RedOrange}{cmyk}{0,0.77,0.87,0} 
\definecolor{Black}{cmyk}{0,0,0,1.} 
\definecolor{ProcessBlue}{cmyk}{0.96,0,0,0} 
\definecolor{NavyBlue}{cmyk}{0.94,0.54,0,0} 
\newcommand\bleu{BleuClairFemto!50}
\newcommand\rouge{RedOrange!50}
\newcommand\gris{gray!60}

\def\scalegrapheres{0.8}
\def\scalegraphesol{0.18}
\def\scalegraphesoltot{0.35}

\let\cite\citep 


\def\volumeyear{2018}
\setcounter{secnumdepth}{3}


\section{Introduction and related work}
\label{sec:intro}
In the context of the decline of fossil fuel resources, the use of
fuel cells appears to be of growing interest as a potential
alternative to conventional power systems~\cite{Jouin2013-IJHE}. Fuel
cells can be used in many applications, such as stationary ones for
domestic use, but also in transportation and portable power
applications~\cite{Borup2007}. Unfortunately, fuel cells suffer from
insufficient durability. Indeed, the current lifetime is usually between
$1\,500$ and $3\,000$ hours, whereas $5\,000$ hours are often required for
transportation applications and up to $100\,000$ hours for stationary
ones. Improving the performance, reliability and lifetime is
therefore an important challenge~\cite{Borup2007} in the fuel cell technology, for which techniques of
Prognostics and Health Management~(\PHM) can definitely help. As recently pointed
out in~\citet{Jouin2013-IJHE}, research in fuel
cells with a PHM viewpoint has mainly focused on data acquisition and data
processing. Less work has been done on condition assessment and
diagnostics and very few papers addressed prognostics and decision
making. Works that take into account the decision part often only proposed
corrective actions (see~\cite{Bosco2000}
and~\cite{Wells2004}), for which physical parameters (such as inlet
and outlet gas flows, pressures and temperatures, single cell and
stacks voltages or current) are controlled in order to master each fuel cell
operating conditions as accurately as possible. These corrective
actions correspond to real-time control (from nanoseconds to seconds),
necessary to compensate the natural fluctuation of fuel cells
parameters and to avoid too early irreversible degradation. At each
time slot the strategy allows to set the operating current to a value that meets the needs in
power for each fuel cell. 

Decision making as addressed in this paper
differs from the previous studies in many aspects. 
In particular, larger time scales (hours
to weeks) are considered and decision comes within the framework of
Prognostic Decision Making (PDM), which aims at optimizing
systems configuration~\cite{Balaban2012}. The system
considered here consists of several fuel cells, used in parallel in order to
provide a global power output. The problem we consider is to provide the power
output value for each fuel cell as a function of time, so as to meet the global power demand. The target application considered here is based on
stationary power generation for domestic usage, also known as micro
combined heat and power (micro-CHP).

In order to deliver suitable power outputs, fuel cells are used in the
form of stacks, composed of many connected single cells. Each
stack is assumed to be independent of the others, but the multi-stack fuel cell
system has to deliver a given global power output. At each time slot, the total provided power output is the sum of
each output of the stacks that are switched on. Each fuel cell
stack is able to deliver an output that can vary continuously within a given interval. The optimization problem consists
in determining the appropriate output for each fuel cell stack during
the whole production horizon. Also the stacks are not constrained to be
all running at each time slot if the target output can be reached by using only
a subset of them. All the stacks may moreover not be available at all times: 
if their end of life has been reached, they are not available for production. Considering a global
power output, the multi-stack system useful life depends not only on
each stack useful life, but also on both the schedule and the
operating condition settings that define the contribution of each
stack over time. The same statement applies to batteries in a health
management context. Saha et al.~\cite{Saha2012} have for instance
addressed the maximization of the battery charge used while
constraining the probability of a battery shut off in flight for
electric unmanned aerial vehicles. Predictions on remaining battery
life are used to optimize mission plans without exceeding the
available battery charge. In a same way, we propose to use prognostics
results in the form of Remaining Useful Lives (\RUL) to maximize the global useful life of a
multi-stack fuel cell system subject to service constraint.

A similar problem has been addressed in~\cite{Herr2014-PHM}
and~\cite{Herr2014-CASE}, where the purpose was to define a schedule
of machines that maximizes the production horizon, based on the
knowledge of each machine remaining useful life (\RUL) in a \PHM\
framework. In these studies, machine throughputs have been assumed to take discrete values. It was shown in~\cite{Herr2014-CASE}
that optimal solutions can be found in reasonable time only for small
size instances and with a very limited number of machines, very few
possible throughput values and short production horizons. An other study
considering this time machines whose performances can vary
continuously between two bounds has been proposed
in~\cite{Herr2017-RenewableEnergy}. The considered model has been built to fit
the fuel cells behavior, but the proposed resolution approach gives
suboptimal solutions and is limited to systems of reasonable size. 

In
order to overcome these two limitations, we propose in this paper to
drastically change the solution paradigm and to build the
solutions globally on the full production horizon. Contribution of
each machine during its lifetime is considered as a whole and
optimized on the full horizon. Each machine contribution is
determined via convex optimization. As a result, our new approach allows to address large problems very quickly.
Concretely, the considered scheduling problem is addressed via optimizing a composite function subject to several  constraints due to fuel cell intrinsic characteristics. A first method is described which performs successive projections onto the sets of constraints. The method generates a sequence of points that can be shown to converge to the solution of the optimization problem~\cite{Bauschke1996}. A second order method is developed and used to define the contribution of each fuel cell stack to the global output over the whole production horizon. It is based on the Mirror Prox method proposed by~\citet{Nemirovski2004} as a variant of the Mirror Descent developed by~\citet{Nemirovsky1983} to minimize a smooth convex function subject to convex constraints. Estimation of the variable is efficient in that it depends very little on its dimension. This is why these methods can be used to solve big optimization problems~\cite{Beck2003}.

The organization of the paper is as follows: the tackled problem is first described in Section~\ref{sec:pbstatement}, with a presentation of the application framework and the optimization problem, followed by a mathematical formulation that complies with the proposed convex resolution paradigm.
The two proposed resolution approaches are then described in Sections~\ref{sec:projections} and~\ref{sec:MPSP}. Efficiency of these methods is assessed through simulation results in Section~\ref{sec:simulationresults}. Conclusion and future work are finally given in Section~\ref{sec:conclusion}.


\section{Problem statement}
\label{sec:pbstatement}

\subsection{Application framework}
\label{sec:applicationframework}
The application addressed in this paper is based on a multi-stack fuel cell system which is supposed to meet energy requirements for domestic usage in a stationary power generation framework. This system is supposed to be composed of $m$ fuel cell stacks $M_j$ ($1\leqslant j\leqslant m$). All the stacks are supposed to be always supplied with raw material required for the energy conversion. They can be used simultaneously and independently from each other.

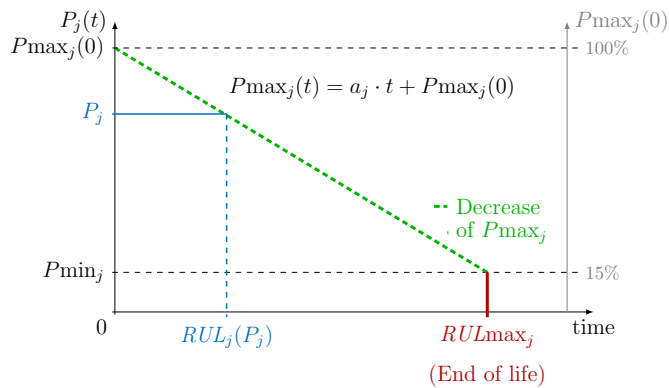
\begin{figure}[htb]
  \center{\scalebox{0.7}{
    \begin{tikzpicture}[scale=0.5]
      \draw[>=latex,->](-0.1,0) -- (18,0) node [below] {time};%
      \draw[>=latex,->](0,-0.1) -- (0,0) node[below left]{$0$} -- (0,11) node[left] {$P_j(t)$};%
      \draw[gray,>=latex,->](17,0) -- (17,11) node[right] {$\Pmax_j(0)$};
      \draw[dashed](-0.1,1.5) node[left]{$\Pmin_j$} -- (17.4,1.5) node[gray,right]{\footnotesize$15\%$};%
      \draw[ultra thick,green!70!black,densely dashed] (0,10) -- (14,1.5);%
      \draw[dashed](-0.1,10) node[left,black]{$\Pmax_j(0)$} -- (17.4,10) node[gray,right]{\footnotesize$100\%$};%
      \node[right] at (4,8.5){$\Pmax_j(t) =  a_j\cdot t + \Pmax_j(0)$};%
      \draw[ultra thick,red!70!black](14,1.5) -- (14,-0.2) node[below] (EOL){$\RULmax_j$};%
      \node[text=red!70!black,below=0 of EOL]{(End of life)};
      \draw[ultra thick,green!70!black,densely dashed](12,4) -- +(0.5,0) node[right]{Decrease};%
      \draw[ultra thick,green!70!black,densely dashed](12.5,3) -- +(0,0) node[right]{of $\Pmax_j$};
      \draw[thick,NavyBlue](-0.1,7.5) node[left]{$P_j$} -- +(4.3,0);%
      \draw[thick,,NavyBlue,dashed](4.2,7.5) -- +(0,-7.7) node[below]{$\RUL_j(P_j)$};%
    \end{tikzpicture}}}
  \caption{Evolution of available power outputs for a machine $M_j$~\label{fig:FCsimplemodele}}
\end{figure}

This corresponds to a parallel machines system, in which each machine is supposed to be able to deliver power outputs $P_j$ that can vary continuously within a given power output range $[\Pmin_j;\Pmax_j]$.
For each machine $M_j$ ($1\leqslant j\leqslant m$), the range of available power outputs depends on the time $t$ and evolves as depicted in Figure~\ref{fig:FCsimplemodele}, with the following extreme output characteristics:
\begin{description}
\item[$\Pmax_j$:] the maximal power output which decreases with time when the machine $M_j$ is used. The straight line depicting its decrease has equation $\Pmax_j(t)=a_j\cdot~t+\Pmax_j(0)$, with $a_j<0$ the speed of the output decline and $\Pmax_j(0)$ the maximal output available at the beginning of the scheduling process. Both parameters of this equation are fuel cells intrinsic characteristics. Useful life of each power output $P_j$, denoted $\RUL_j(P_j)$, is limited by this maximal power output decrease ;
\item[$\Pmin_j$:] the minimal power output, constant over time and associated to the maximal Remaining Useful Life: $\RUL_j(\Pmin_j) = \max_{\Pmin_j\leqslant P_j\leqslant \Pmax_j}(\RUL_j(P_j)) = \RULmax_j$.
\end{description}

The model proposed in Figure~\ref{fig:FCsimplemodele} is a simplified version of the one developed in~\cite{Herr2017-RenewableEnergy}.
It considers however the main characteristics of fuel cells useful for the scheduling purpose considered here. Due to the decrease over time of the maximal power output $\Pmax_j(t)$, this model is asymmetric. In a schedule defining the contribution over time of a machine, two periods of time in which a machine is used with different outputs can then not always be permuted (see~\cite{Herr2017-RenewableEnergy} for an illustration of this property).

\subsection{Optimization problem}
At each time $t$ ($0\leqslant t\leqslant T$), the global outcome $\Ptot$ is the sum of each stack power contribution. During the whole production horizon,
this global outcome has to reach a given load demand $\sigma(t)$.
Overproduction is authorized if it allows to extend the global system useful life (or production horizon), but should be avoided as far as possible. In fact, storage being not considered in this study, overproduction is supposed to be lost. All the machines are not supposed to be in use at each time if a subset of them is enough to reach the demand. Since machines suffer from wear and tear, some machines can also be unavailable at a certain time if their end of life~(\EOL) has been reached. Stop-and-start of fuel cell stacks have moreover to be avoided as far as possible. Stopping and restarting a fuel cell can indeed induce considerable damage and lead to premature aging~\cite{Borup2007}. Change of power output during the use of fuel cell stacks is however still authorized.

Considering these assumptions, the point is to manage the system by defining the commitment of fuel cell stacks so as to reach the demand as long as possible. During the whole production horizon, the purpose is then to define at each time each stack contribution to the global power output.

\subsection{Mathematical formulation}
\label{sec:mathformulation}
In this section, a mathematical formulation of the problem previously described is defined, making use of convex elements. Let $f_j(t)$ ($1\leqslant j\leqslant m$, $0\leqslant t\leqslant T$)
be the vector defining the evolution over time of the power output delivered by the machine $M_j$, with $T$ the length of the decision horizon. Link between this decision horizon and the solution production horizon, denoted $H$, is clarified in Section~\ref{sec:postprocessing}. Contributions of all the machines are gathered together in a vector $F\in \mathbb{R}^{m(T+1)}$ such that:
\[F = \left[f_{1}(0),\ldots, f_{1}(T), \ldots, f_j(t),\ldots f_{m}(0), \ldots, f_{m}(T)\right].\]

The general idea is to
minimize a convex function subject to a set of constraints. The
objective function aims at ensuring that the power demand is
reached. At each time $t$ ($0\leqslant t\leqslant T$), it is about
minimizing the difference between the global power output delivered by
the set of machines and the demand $\sigma(t)$.
This is expressed by Equation~\eqref{eq:objfct}, where $\phi$ measures the error incurred by the choice of $f_j$, $\forall 1\leqslant j\leqslant m$. One possible choice of $\phi$ is the squared loss, i.e. the squared $L_2$ norm.
\begin{equation}
  \label{eq:objfct}
  \min \quad \phi\left(\sigma(\cdot) - \sum\limits_{j=1}^m{f_j(\cdot)}\right) 
\end{equation}

Constraints on each function $f_j$ relate to the definition domain of each contribution and to the limited availability of machines.
At each time $t$, 
each machine contribution is either equal to zero or constrained
between two bounds (see Equation~\eqref{eq:bounds}), in accordance
with the 
hypotheses detailed in the application framework.
$f_j(t)=0$ means that the machine $M_j$ is not used at time $t$.
\begin{equation}
  \label{eq:bounds}
  f_j(t) = 0 \quad \text{or} \quad \fmin_j(t)\leqslant f_j(t)\leqslant \fmax_j(t)
\end{equation}
\[\forall\, 1\leqslant~j\leqslant~m,\, \forall\, 0\leqslant~t\leqslant~T\]
Each contribution $f_j$ is constrained by the maximal power output
decrease of the associated machine $M_j$, which expresses its limited
availability. Evolution over time of this maximal power output,
$\fmax_j(t)$, is a function of the use of machine $M_j$,
$f_j(t)$. Indeed, $\fmax_j(t)$ evolves only if $M_j$ is used, that is,
if $f_j(t)>0$. A first formulation is proposed in
Equation~\eqref{eq:evolfmax}, with $a_j$ the speed associated to the
maximal power output decrease ($a_j<0$).
\begin{equation}
  \label{eq:evolfmax}
  \displaystyle \fmax_j(t) = \left\{
    \begin{array}{ll}
      \fmax_j(t-1) + a_j & \text{if } f_j(t) > 0\, ;\\
      \fmax_j(t-1) & \text{if } f_j(t) = 0
    \end{array}
  \right.  
\end{equation}
\[\forall\, 1\leqslant~j\leqslant~m,\, \forall\, 1\leqslant~t\leqslant~T\]

\begin{figure}[h]
  \center{\scalebox{0.7}{
    \begin{tikzpicture}[scale=0.5]
      \draw[>=latex,->](-0.1,0) -- (18,0) node [below] {time};%
      \draw[>=latex,->](0,-0.1) -- (0,0) node[below left]{$0$} -- (0,11) node[left] {$P_j(t)$};%
      \draw[gray,>=latex,->](17,0) -- (17,11) node[right] {$\Pmax_j(0)$};
      \draw[dashed](-0.1,1.5) node[left]{$\Pmin_j$} -- (17.4,1.5) node[gray,right]{\footnotesize$15\%$};%
      \draw[ultra thick,green!70!black,densely dashed] (0,10) -- (14,0);%
      \draw[dashed](-0.1,10) node[left,black]{$\Pmax_j(0)$} -- (17.4,10) node[gray,right]{\footnotesize$100\%$};%
      \node[right] at (3.5,8.5){\footnotesize$\fmax_j(t) = \fmax_j(t-1) + a_j$};
      \draw[ultra thick,red!70!black](14,0.2) -- (14,-0.2) node[below] (EOL){$\RULmax_j$};%
      \draw[ultra thick,green!70!black,densely dashed](12,4) -- +(0.5,0) node[right]{Decrease};%
      \draw[ultra thick,green!70!black,densely dashed](12.5,3) -- +(0,0) node[right]{of $\Pmax_j$};
    \end{tikzpicture}}}
  \caption{Simplified convex evolution of available power outputs for a machine $M_j$\label{fig:FCmodeleconvexe}} 
\end{figure}
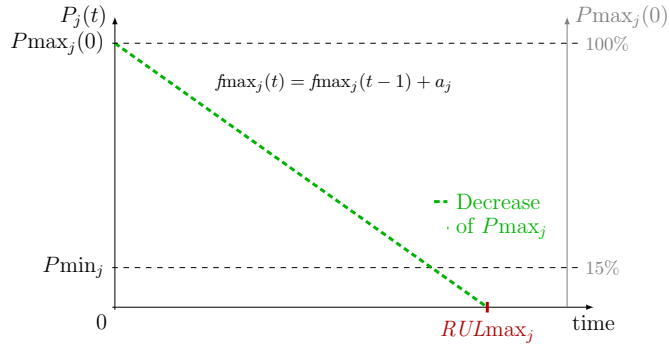

Equations~\eqref{eq:bounds} and~\eqref{eq:evolfmax} being not convex, they can not be used as is within the proposed convex programming paradigm.
A second formulation of the constraints is proposed in set of equations~\eqref{eq:pgconvexe}, which details the mathematical program associated to the optimization problem. This program does not respect the real evolution over time of the maximal power output that can be delivered by machines,
but presents the advantages of being convex and thus consistent with the convex resolution methods proposed in next section. In the following, the machines behavior follows the simplified model depicted in Figure~\ref{fig:FCmodeleconvexe}.
\begin{subnumcases}{\hspace*{-0.8cm}\label{eq:pgconvexe}}
\footnotesize \hspace*{-0cm}
  \strut \min \quad \phi\left(\sigma(\cdot) - \sum\limits_{j=1}^m{f_j(\cdot)}\right)\label{eq:pgconvobj}\\
  \hspace*{-0cm}\, \tq \quad \fmax_j(t)\leqslant \fmax_j(t-1)\nonumber\\
  \hspace*{-0cm}\quad\quad\quad\quad + \mu\cdot a_j\cdot(f_j(t-1))^{\upsilon} \quad \forall\, 1\leqslant j\leqslant m, \label{eq:pgconvfmax}\\
  \hspace*{-0cm}
  \quad\quad\quad\quad\quad \forall 1\leqslant t\leqslant T, \text{with } \mu \text{ and } \upsilon\in [0,1]\nonumber\\
  \hspace*{-0cm}\, \text{with} \quad 0\leqslant f_j(t)\leqslant \fmax_j(t)\label{eq:pgconvbornes}\\
  \hspace*{-0cm}\quad\quad\quad\quad \forall 1\leqslant j\leqslant m,\, \forall 0\leqslant
  t\leqslant T\nonumber
\end{subnumcases}


\section{Resolution based on successive projections onto the sets of constraints} 
\label{sec:projections}
The first proposed resolution approach makes use of convex projections to cope with the scheduling problem. The general idea is to perform successive projections onto each set of constraints previously defined in the convex program~\eqref{eq:pgconvexe} in order to generate a sequence of points that converges to the solution of the considered optimization problem. The three projections are first detailed. The global projection scheme is then described.

\subsection{Projections}
\label{sec:detailedproj}
Each projection allows to comply with one of the three main constraints defined by Equations~\eqref{eq:pgconvobj}, \eqref{eq:pgconvfmax} and~\eqref{eq:pgconvbornes}.

\subsubsection{Reaching of the demand}
\label{sec:projdemand}
First constraint corresponds to the reaching of the demand $\sigma(t)$ at each time $t$ ($0\leqslant t\leqslant T$). The decision horizon is split into time intervals of length $\Delta{T}$, with $1\leqslant \Delta{T}\leqslant T$. For each time interval, delimited by the times $\tstart$ and $\tend$ such that $\tstart\leqslant t\leqslant \tend$, the projection of the solution $F$ on the demand $\sigma$ is performed following the strategy detailed hereafter:
\begin{description}
\item[Machine selection:] machines $M_j$ ($1\leqslant j\leqslant m$) are sorted following an ascending order of the difference $(\fmax_j(t) - f_j(t))$. This sorting favors machines whose contributions before the projection are the closest from their maximal reachable output $\fmax_j$.
This allows to use first machines already started and to avoid stop-and-start of fuel cells, which has been shown to lead to premature aging~\cite{Borup2007}.

With $\distfmax(j) = \max(\fmax_j(\tend) - f_j(\tend), 0)$ being the difference between the current contribution of machine $M_j$ and its maximal reachable contribution, the subscript of the machine $M_{j'}$ whose contribution has to be modified first is determined by Equation~\eqref{eq:machselect}.
\begin{equation}
\label{eq:machselect}
j' = \argmin_{1\leqslant j\leqslant m | \distfmax(j)>0}{\distfmax(j)}
\end{equation}

\item[Projection method:] the contribution of the selected machine $M_{j'}$ is increased on each time interval of length $\Delta{T}$ just enough to reach the demand $\sigma$. The increase being $\inc = \sigma(\tend) - \sum_{j=1}^m{f_j(\tend)}$, the contribution of machine $M_{j'}$ after the projection is defined by Equation~\eqref{eq:machproj}.
\begin{align}
\label{eq:machproj}
f_{j'}(t) = & \min{(f_{j'}(\tend) + \inc, \fmax_{j'}(\tend))}\\
& \forall \tstart\leqslant t\leqslant \tend\nonumber
\end{align}
\end{description}

The selection of a machine and the projection of its contribution on the demand are performed on each time interval of length $\Delta{T}$, until the demand is reached or until there is no machine available anymore.

\subsubsection{Evolution of the maximal output} Each maximal output reachable $\fmax_j$ is updated as a function of the contribution of the corresponding machine, $f_j$, following the relation defined in Equation~\eqref{eq:projfmax}.
\begin{align}
\label{eq:projfmax}
\fmax_j(t) & = \fmax_j(t-1) + \mu\cdot a_j\cdot(f_j(t-1))^{\upsilon}\\
& \forall\, 1\leqslant j\leqslant m, \forall 1\leqslant t\leqslant T, \text{with } \mu \text{ and } \upsilon\in [0,1]\nonumber
\end{align}

\subsubsection{Respect of the maximal output} A projection of each machine contribution $f_j$ onto the corresponding $\fmax_j$ can be necessary to fix some possible overrun and to comply with the updated maximal reachable output. This projection simply limits each contribution $f_j(t)$ to the corresponding $\fmax_j(t)$ defined by the application of the previous projection.
\begin{align}
\label{eq:projlimitfmax}
f_j(t) & = \fmax_j(t) \quad \text{if } f_j(t)>\fmax_j(t)\\
& \forall 1\leqslant j\leqslant m,\, \forall 0\leqslant t\leqslant T\nonumber
\end{align}

\subsection{Resolution algorithm based on the projections}
\label{sec:algoproj}
The proposed resolution scheme
performs successive projections on the three main constraints, following the strategies previously defined.
The positivity of each component of the solution vector $F$ is ensured through a positive initialization of each machine contribution at each time ($f_j^{(0)}(t)\geqslant 0\, \forall 1\leqslant j\leqslant m, \forall 0\leqslant t\leqslant T$).


The evolution of the vectors $\fmax_j$ being determined as a function of each $f_j$, a combined determination of these two elements is required and several successive launches of all the projections are necessary to optimize the solution. The definition of a schedule that maximizes the production horizon while complying with all the constraints requires then several iterations of the sequence composed of the three successive projections.
This sequence is iterated until a stopping condition is verified. The number of iterations necessary to obtain a satisfying solution depends indeed strongly on the configuration of the considered problem, that is, on the number of machines, on the shape of the maximal reachable outputs $\fmax_j$ and on the demand $\sigma$. The optimization process of the solution $F$ is then stopped when it does not change significantly anymore, that is, as soon as the difference between two successive values of F ($|F - F^{prev}|$) gets below a certain threshold $\epsilon$.

Solutions obtained with this first resolution method are piecewise defined and the reached production horizons depend strongly on the initialization of the solution vector $F$ and on the projection strategy. This first resolution approach does then not fully comply with the resolution paradigm proposed in this paper, which aims to build the solutions globally on the full production horizon. It allows however to obtain solutions very quickly, which can serve as a point of comparison with the solutions obtained with the resolution method proposed in next section.

\section{Resolution based on a smooth penalization approach using the Mirror Prox}
\label{sec:MPSP}
The second resolution method is based on a smooth penalisation approach. A Mirror Prox method is proposed to be used to cope with the problem of minimizing the objective function detailed previously in Equation~\eqref{eq:pgconvobj}.
\subsection{Principle of the Mirror-Prox}
The Mirror Prox algorithm, developed by~\citet{Nemirovski2004}, is a variant of the Mirror Descent algorithm, which has first been proposed by~\citet{Nemirovsky1983} for convex programming. It has been extensively studied recently and several relationships have been discovered between the Mirror Descent scheme and Bregman-proximal methods.

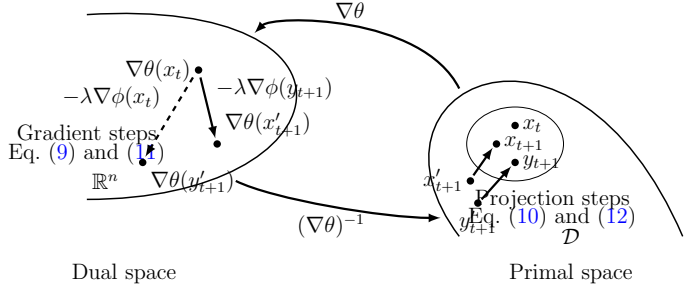
\begin{figure}[htb]
  \hspace*{-0.3cm} \scalebox{0.7}{
    \begin{tikzpicture}[scale=0.7]
      \draw[thick] (0,6) .. controls (8,6.5) and (7,1) .. (0,1);
      \node at (0,2.75){Gradient steps};
      \node at (0,2.25){Eq.~\eqref{eq:MP1} and~\eqref{eq:MP3}};
      \node at (0.5,1.5){$\mathbb{R}^n$};
      \node at (1,-1){Dual space};
      \fill (3,4.5) circle (0.1) node[left]{$\nabla \theta(x_t)$};
      \fill (3.5,2.5) circle (0.1) node[above right]{$\nabla \theta(x'_{t+1})$};
      \draw[>=latex,->,very thick] (3.05,4.3) --(3.45,2.65)
      node[midway,above right]{$- \lambda \nabla \phi(y_{t+1})$};
      \fill (1.5,2) circle (0.1) node[below right]{$\nabla \theta(y'_{t+1})$};
      \draw[>=latex,->,very thick,dashed] (2.85,4.3) --(1.57,2.15)
      node[midway,above left]{$- \lambda \nabla \phi(x_t)$};

      \draw[thick] (10,0) .. controls (7,4) and (13,7) .. (16,0);
      \node at (12.5,1){Projection steps};
      \node at (12.5,0.5){Eq.~\eqref{eq:MP2} and~\eqref{eq:MP4}};
      \node at (13,0){$\mathcal{D}$};
      \node at (13,-1){Primal space};
      \fill (10.5,0.9) circle (0.1) node[below]{$y'_{t+1}$};
      \fill (11.5,2) circle (0.1) node[right]{$y_{t+1}$};
      \fill (11,2.5) circle (0.1) node[right]{$x_{t+1}$};
      \fill (11.5,3) circle (0.1) node[right]{$x_{t}$};
      \fill (10.3,1.5) circle (0.1) node[left]{$x'_{t+1}$};
      \draw (11.5,2.5) ellipse (1.3 and 1);
      \draw[>=latex,->,very thick] (10.6,1) --(11.4,1.9);
      \draw[>=latex,->,very thick] (10.38,1.65) --(10.9,2.4);

      \draw[>=latex,->,very thick] (10,4) .. controls (9,6) and
      (5,6) .. (4.5,5.5) node[midway,above]{$\nabla \theta$};
      \draw[>=latex,->,very thick] (4,1.5) .. controls (5,1) and
      (8,0.5) .. (9.5,0.5) node[midway,below]{$(\nabla \theta)^{-1}$};
    \end{tikzpicture}}
  \caption{Operating principle of the Mirror Prox method, based on~\cite{Bubeck2014}\label{fig:MirrorProx}}
\end{figure}

Both approaches are based on the resolution of a primal-dual saddle point problem, which allows to take constraints into account. The purpose is to minimize a smooth convex function
under constraints. The Mirror Descent algorithm makes use of a gradient descent to find the minimum of the considered function. The goal is to minimize the local linearization of the function while not moving too far away from the previous point, with distances measured via the Bregman divergence of the mirror map. A mirror function allows to transition from the primal space, where all the constraints of the problem are defined, to the dual space. The Mirror Prox method applies at each iteration two consecutive steps of Mirror Descent. A very instructive description of the Mirror Prox algorithm has been proposed by~\citet{Bubeck2014}.
As illustrated in Figure~\ref{fig:MirrorProx}, the Mirror Prox algorithm applies a first time the Mirror Descent to go from $x_t$ to $y_{t+1}$ and then a second similar step to obtain $x_{t+1}$. This second step stats again from $x_t$, but uses the gradient of $f$ evaluated at $y_{t+1}$ (instead of $x_t$). These steps are defined by Equations~\eqref{eq:MP1} to~\eqref{eq:MP4}.
\begin{equation}
  \label{eq:MP1}
  \nabla \theta(y'_{t+1}) = \nabla \theta(x_t) - \lambda \nabla \phi(x_t) \quad \quad \text{with } \lambda\geqslant 0
\end{equation}
\begin{equation}
  \label{eq:MP2}
  y_{t+1} = P_{\mathcal{C}}(\nabla \theta(y'_{t+1})) = \argmin_{x\in \mathcal{C}\cap\mathcal{D}}{D_{\theta}(x,y'_{t+1})} 
\end{equation}
\begin{equation}
  \label{eq:MP3}
  \nabla \theta(x'_{t+1}) = \nabla \theta(x_t) - \lambda \nabla \phi(y_{t+1}) \quad \quad \text{with } \lambda\geqslant 0
\end{equation}
\begin{equation}
  \label{eq:MP4}
  x_{t+1} = P_{\mathcal{C}}(\nabla \theta(x'_{t+1})) = \argmin_{x\in \mathcal{C}\cap\mathcal{D}}{D_{\theta}(x,x'_{t+1})} 
\end{equation}

\subsection{Resolution algorithm based on the Mirror Prox}
The proposed resolution algorithm makes use of the Mirror Prox scheme previously described. The resolution of the primal-dual problem being highly sensitive to the variation of the different parameters associated to the model, the proposed formulation incorporates the different constraints of the application directly into the objective function. The considered objective function $\phi(F,\fmax,\sigma)$, detailed in Equation~\eqref{eq:fctobjMPSP}, aims then at satisfying both the reaching of the demand and the respect of the evolution of the maximal outputs $\fmax_j(t)$, $\forall 1\leqslant j\leqslant m$ and $\forall 0\leqslant t\leqslant T$.
\begin{align}
  \label{eq:fctobjMPSP}
  \phi(F,\fmax,\sigma) = & \lambda_{dem} h_{dem}(F,\sigma) +
  \lambda_{slope} h_{slope}(F,\fmax)\nonumber\\
  & \quad\quad\quad \text{with } \lambda_{dem},\, \lambda_{slope} > 0
\end{align}

The first component of this objective function, $h_{dem}(F,\sigma)$, detailed in Equation~\eqref{eq:hdem}, aims at satisfying the demand $\sigma(t)$ at each time $t$. We did not choose the $L_2$ loss because asymetric loss are slightly more relevant in our context. In fact, finding a solution which does not match the demand should be more heavily penalised than a solution which matches the demand by far. This cost function, based on the objective function previously defined in Equation~\eqref{eq:pgconvobj}, allows to heavily penalize $F$ when the demand becomes larger than the total output provided by the set of machines and conversely to penalize it only slightly when the production is larger than the demand. Second part of the objective function, $h_{slope}(F,\fmax)$, is used to control the evolution of each $\fmax_j$ as a function of each associated contribution $f_j$ (see equation~\eqref{eq:hslope}). This corresponds to the penalization associated to the constraint previously defined by Equation~\eqref{eq:pgconvfmax}.
\begin{align}
  \label{eq:hdem}
  h_{dem}(F,\sigma) = & \sum\limits_{t=0}^T{\frac{1}{T+1} \exp{\bigg(-\gamma\bigg(\sum\limits_{j=1}^m{f_j(t) -
        \sigma(t)}\bigg)\bigg)}}\nonumber\\
        & \quad\quad\quad \text{with } \gamma > 0
\end{align}
\begin{align}
  \label{eq:hslope}
  h_{slope}(F,\fmax) = & \sum\limits_{t=1}^T{\sum\limits_{j=1}^m{\exp{\bigg(\delta \bigg(\fmax_j(t) - \fmax_j(t-1)}}} \nonumber\\
  & - \mu' a_j(f_j(t-1))^{\upsilon'}\bigg)\bigg)\\
  & \text{with } \delta > 0,\, \mu' > 0,\, \upsilon' > 1\nonumber
\end{align}

The mirror function $\theta$ considered in the Mirror Prox formula is defined on $\mathbb{R}^{m(T+1)}$ by Equation~\eqref{eq:mirrormap} and its gradient, which is used in the Mirror Descent steps, is expressed in Equation~\eqref{eq:gradmirrormap}.
\begin{align}
  \label{eq:mirrormap}
  \theta(F) = \sum\limits_{t=0}^T \sum\limits_{j=1}^m{F \ln(F)} 
\end{align}
\begin{align}
  \label{eq:gradmirrormap}
  \nabla \theta(F) = \ln(F) + 1
\end{align}

The Mirror Prox used in the proposed resolution algorithm is then of the following form (Equations~\eqref{eq:MPint1} to~\eqref{eq:MPint4}):
\begin{align}
  \label{eq:MPint1}
  \nabla \theta(F^{(l+1)}) = & \nabla \theta(F^{(l)}) - \lambda \nabla_F \phi(F^{(l)},\fmax,\sigma)\\
  & \quad\quad\quad \text{with } \lambda\geqslant 0\nonumber
\end{align}
\begin{align}
  \label{eq:MPint2}
  F_{int} = \exp\left(\nabla \theta(F^{(l+1)}) - 1\right)
\end{align}
\begin{align}
  \label{eq:MPint3}
  \nabla \theta(F^{(l+1)}) = & \nabla \theta(F^{(l)}) - \lambda \nabla_{F_{int}} \phi(F^{(l)},\fmax,\sigma)\\
  & \quad\quad\quad \text{with } \lambda\geqslant 0\nonumber
\end{align}
\begin{align}
  \label{eq:MPint4}
  F^{(l+1)} = \exp\left(\nabla \theta(F^{(l+1)}) - 1\right)
\end{align}
with the gradient of $\phi$ defined by Equations~\eqref{eq:gphi}
to~\eqref{eq:ghslope}.
\begin{align}
  \label{eq:gphi}
  \nabla_F \phi(F^{(l)},\fmax,\sigma) = & \lambda_{dem} \nabla_F{h_{dem}(F,\sigma)}\nonumber\\
  & + \lambda_{slope} \nabla_F{h_{slope}(F,\fmax)}\\
  & \quad\quad\quad \text{with } \lambda_{dem},\, \lambda_{slope} > 0\nonumber
\end{align}
\begin{align}
  \label{eq:ghdem}
  \nabla_F{h_{dem}(F,\sigma)} = & -\gamma \sum\limits_{t=0}^T \frac{1}{T+1}\cdot\nonumber\\
  & \exp{\bigg(-\gamma\bigg(\sum\limits_{j=1}^m{f_j(t) - \sigma(t)}\bigg)\bigg)}\\
  & \quad\quad\quad \text{with } \gamma > 0\nonumber
\end{align}
\begin{align}
  \label{eq:ghslope}
  \nabla_F & {h_{slope}(F,\fmax)} = -\delta \mu' \upsilon'
  \sum\limits_{t=0}^T\sum\limits_{j=1}^m a_j\cdot f_j(t-1)^{\upsilon'-1}\cdot\nonumber\\
  & \exp{\bigg(\delta \Big(\fmax_j(t) - \fmax_j(t-1) - \mu' a_j(f_j(t-1))^{\upsilon'}}\Big)\bigg)\nonumber\\
  & \quad\quad\quad\quad\quad\quad \text{with } \delta > 0,\, \mu'>0,\, \upsilon' > 1
\end{align}

In order to accelerate the convergence of the method, the distance of the solution to its fictional projection onto the demand $\sigma(t)$, $(F - F^{proj})$, is proposed to be added to each gradient step performed by the Mirror Descent steps. This allows to accelerate the evolution of the solution towards a state which complies at best with the different constraints. The new formulation of the Mirror Prox is then detailed in Equations~\eqref{eq:MPfin1} to~\eqref{eq:MPfin4}, with the vector $F^{proj}$ determined by a projection of the solution vector $F$ onto the demand $\sigma(t)$,
with $gF^{proj} = w_{grad} (F^{(l)} - F^{proj})$ and $w_{grad} > 0$.
\begin{align}
  \label{eq:MPfin1}
  \nabla \theta(F^{(l+1)}) = & \nabla \theta(F^{(l)})\nonumber\\
  & - \lambda \left(\nabla_F \phi(F^{(l)},\fmax,\sigma) + gF^{proj}\right)\\
  & \quad\quad\quad \text{with }\lambda\geqslant 0\nonumber
\end{align}
\begin{align}
  \label{eq:MPfin2}
  F_{int} = \exp\left(\nabla \theta(F^{(l+1)}) - 1\right)
\end{align}
\begin{align}
  \label{eq:MPfin3}
  \nabla \theta(F^{(l+1)}) = & \nabla \theta(F^{(l)})\nonumber\\
  & - \lambda \left(\nabla_{F_{int}} \phi(F^{(l)},\fmax,\sigma) +
    gF^{proj}\right)\nonumber\\
    & \quad\quad\quad \text{with } \lambda\geqslant 0
\end{align}
\begin{align}
  \label{eq:MPfin4}
  F^{(l+1)} = \exp\left(\nabla \theta(F^{(l+1)}) - 1\right)
\end{align}

The use of this additional gradient step allows furthermore to define a stopping condition for the global process,
which allows to avoid the definition of a number of iterations adapted to the method and to the optimization problem parameters. The resolution process is stopped when the differences between all the components of the solution vector $F$ and the corresponding components of the vector $F^{proj}$ are all lower than a certain threshold. In other words, the resolution process is stopped when the reaching of the demand can not be improved anymore. After each step of the Mirror Prox, several projections onto the constraints are moreover performed in order to guarantee the respect of these constraints and to improve the convergence speed of the solution. An update of the maximal power outputs reachable, $\fmax_j$, as a function of each vector $f_j$, allows to accelerate their evolution even if their progression is handled by the Mirror Prox. A projection of the machines contributions $f_j$ onto the corresponding $\fmax_j$ allows finally to comply with the maximal power output limitation.


\section{Simulation results}
\label{sec:simulationresults}

\subsection{Problem generation}
Random problem configurations have been generated using a simulator and configured with many parameters including the number of machines in the considered multi-stack fuel cell system, $m$, and intrinsic fuel cell characteristics. The latter have been defined on the basis of fuel cell manufacturer specifications and considering a maximal lifetime $\RULmax_j=1500$ hours $\pm 20\%$ for each machine $M_j$ ($1\leqslant j\leqslant m$). Each $\RULmax_j$ value is drawn following a uniform distribution between $1200$ and $1800$ hours. Power output values are determined in the same way, with $\Pmax_j(0)=500\, W\, \pm 5\%$ and $\Pmin_j=0.15\cdot \Pmax_j(0)$ for each machine $M_j$.

For the results presented hereafter, the power demand has been assumed to be constant during the whole scheduling horizon: $\sigma(t) = \sigma$ $\forall 0\leqslant t\leqslant T$. Without any loss of generality, only one demand value has then been associated to each problem configuration, but many demands corresponding to different configurations have been tested. Many loads $\alpha$ have been defined such that $\sigma=\alpha\cdot\Pnom_{tot}$, with $\Pnom_{tot}$ the nominal total power output reachable with the considered fuel cells system and $30\% \leqslant \alpha \leqslant 90\%$. $\Pnom_{tot} = \sum\nolimits_{j=1}^m{\Pnom_j}$, with $\Pnom_j=0.75\cdot\Pmax_j(0)$ the power output recommended by fuel cell manufacturers for a nominal use of fuel cells. In the following figure, results are represented as a function of the load $\alpha$.

\subsection{Resolution methods configuration}
\label{sec:config}
Initial values of each solution vector $F$ has first been set to zero: $f_j(t)=0$ $\forall\, 1\leqslant j\leqslant m$, $\forall\, 0\leqslant t\leqslant T$. Quality of solutions from the point of view of the reached production horizon globally increases with the number of iterations and stabilizes starting from a certain value. For each resolution method proposed earlier, several iterations of the associated process are then performed to optimize this solution. The global process is stopped when the variation of the solution vector is not significant anymore, that is when $(F~-~F^{proj})~<~\epsilon$ for the Mirror-Prox-based algorithm (resp. $(F~-~F^{prev})~<~\epsilon$ for the algorithm based on the successive projections), with the threshold $\epsilon$ defined as a function of the demand value as follows: $\epsilon = 0.1\cdot\sigma$.

Tuning of the different parameters involved in the two resolution methods allows to comply with the constraints and to adapt the shape of each $\fmax_j$. For the evolution of the maximal power output $\fmax$, values have been defined as follows: $\mu=0.2$, $\upsilon=0.3$. Values for the parameters used in the Mirror-Prox algorithm are the following: $\lambda=5\cdot 10^{-5}$, $\lambda_{dem}=\lambda_{slope}=100$, $\gamma=100$, $\delta=100$, $\mu'=1$, $\upsilon'=1$.

\subsection{Post-processing}
\label{sec:postprocessing}
The main constraint of the optimization problem is the reaching of the power output demand $\sigma(t)$. This constraint being tackled through the minimization of an objective function, solutions may contain time periods during which this demand is not reached. But, with the two resolution methods proposed in previous section, solutions are built so that the time periods for which the power demand $\sigma(t)$ is reached are gathered at the beginning of the schedules. This is consistent with the objective to maximize the production horizon of the set of machines.
This behavior is linked with the shape of the functions $\fmax_j(t)$ representing for each machine $M_j$ the evolution over time of the maximal power output reachable, which limits the contribution of each machine. These functions being strictly decreasing with the use of machines, it is in fact more likely to reach the demand at the beginning of the scheduling process than after some time. As already mentioned, resolution algorithms detailed previously can then be applied on overestimated horizons $T$, named decision horizons. The production horizon of each solution, $H$, is simply the maximal time during which all the constraints are strictly satisfied. In practice, the production horizon corresponds to the time during which the demand $\sigma(t)$ is reached.


\subsection{Results}
Efficiency of the proposed commitment strategies defined in Sections~\ref{sec:projections} and~\ref{sec:MPSP} is assessed through the comparison of reached production horizons to a theoretical upper bound. Considering a constant demand $\sigma$ and a set of fuel cell stacks, this upper bound, denoted \UB\ and defined in Equation~\eqref{eq:BScontinu}, corresponds to the theoretical maximal time during which the demand can be reached.
\begin{equation}
  \label{eq:BScontinu}
  \UB = \left\lfloor
    \frac{\sum\limits_{j=1}^{m}{0.6\cdot\Pmax_j(0)\cdot\RUL_j(\Pmin_j)}}{\sigma}\right\rfloor
\end{equation}

\begin{figure}[htb]
  \centerline{\includegraphics[width=\linewidth]{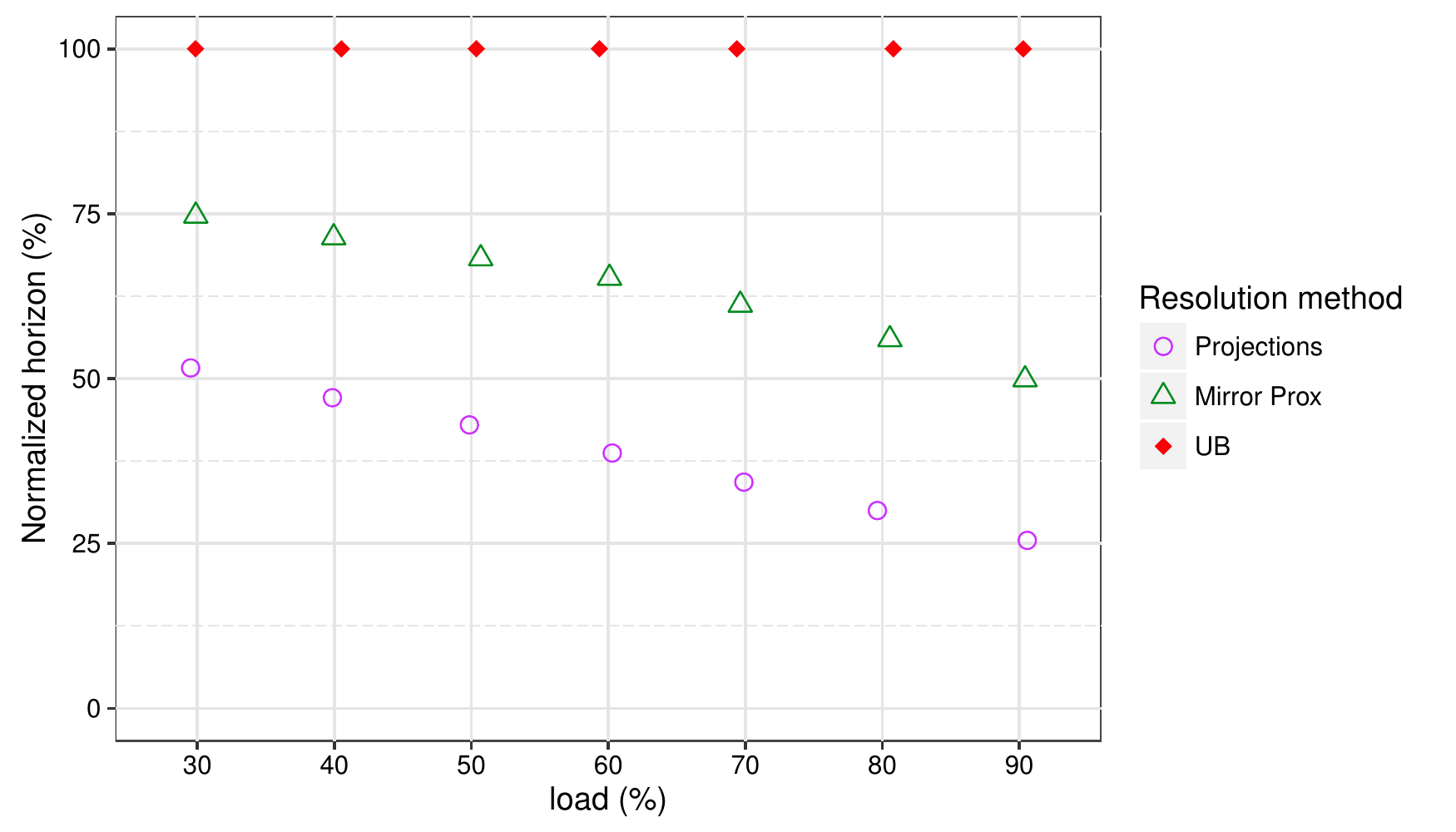}}
  \caption{Normalized horizon -- m = 25 machines} 
  \label{fig:DistUBConvex}
\end{figure}

Figure~\ref{fig:DistUBConvex} shows the production horizons obtained with the considered strategies normalized with the upper bound, when considering a set of $25$ fuel cell stacks. This upper bound being never reachable, results are actually better than showed. One can see that the first strategy, performing successive convex projections, allows to reach a mean relative horizon of around $39\%$ of the upper bound \UB\ and does not exceed $51,6\%$. The resolution method based on the Mirror Prox gives better results. It allows indeed to reach a mean relative horizon of around $64.3\%$ of \UB\ and $74,7\%$ for the best case.

\begin{figure}[htb] \centering
  \begin{subfigure}[b]{\linewidth}
    \centerline{\includegraphics[scale=\scalegraphesol]{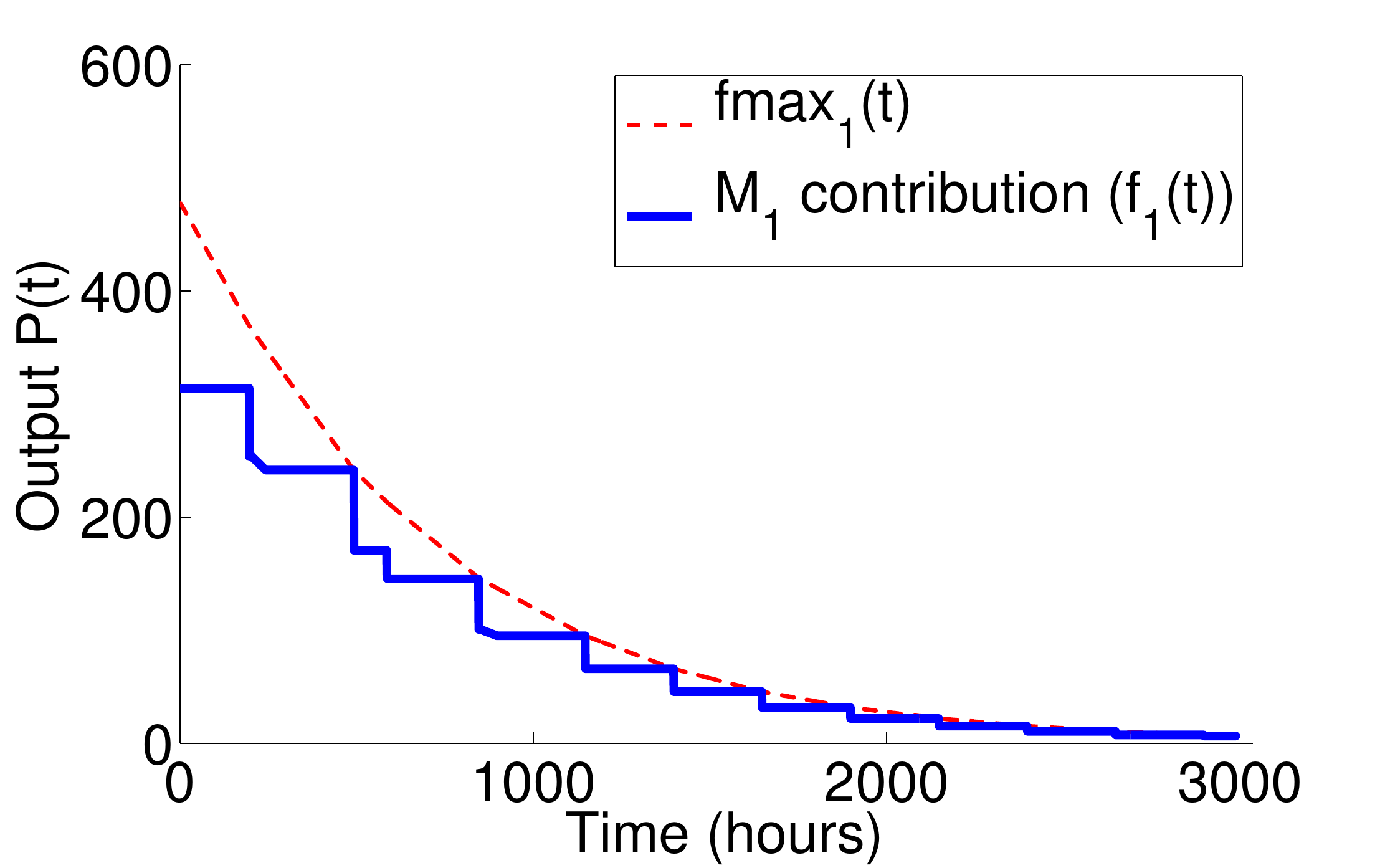}
    \includegraphics[scale=\scalegraphesol]{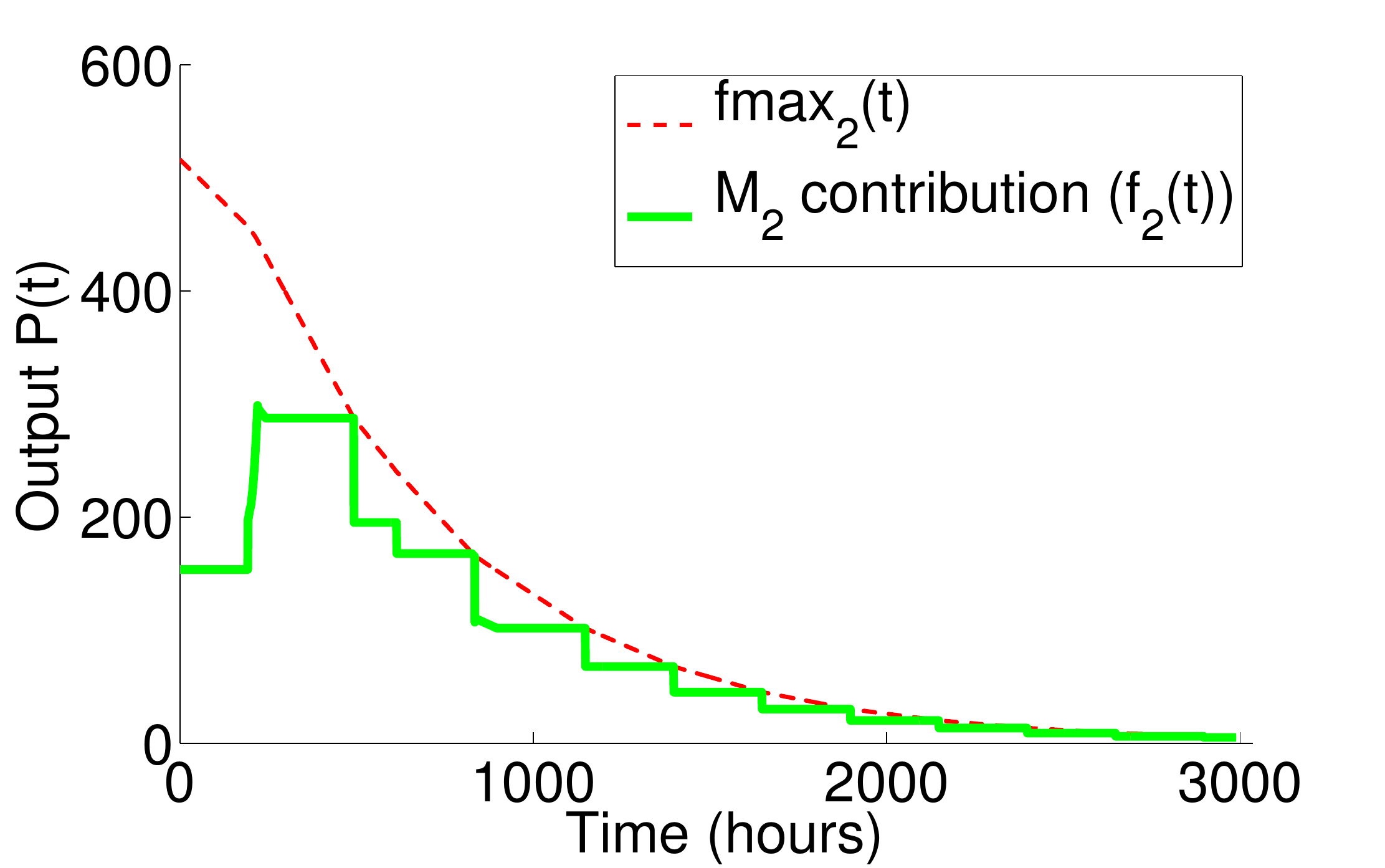}}

    \centerline{\includegraphics[scale=\scalegraphesol]{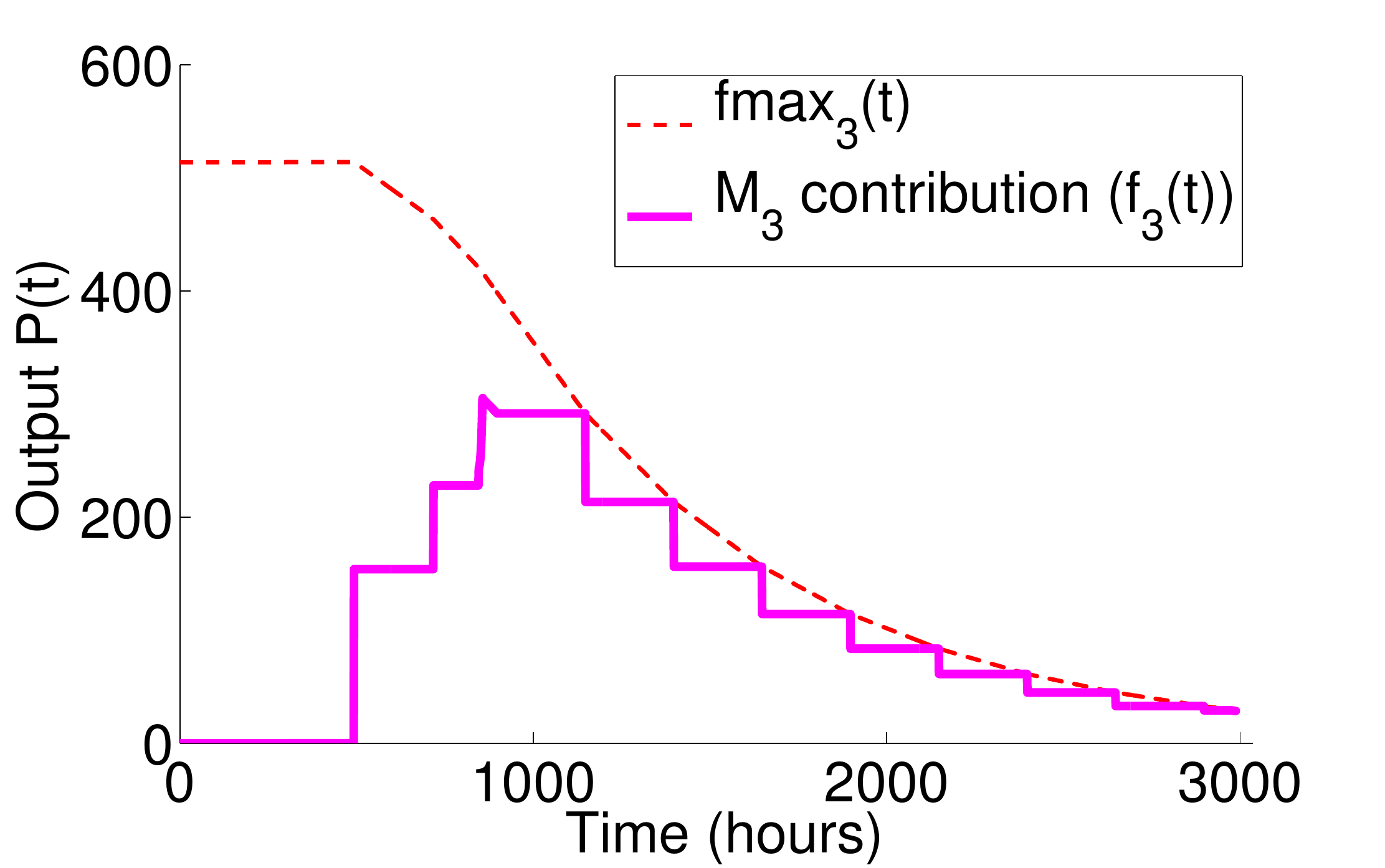}}
    \caption{Machines contributions\label{fig:detailcontributionsProj}}
  \end{subfigure}

  \begin{subfigure}[b]{\linewidth}
    \centerline{\includegraphics[width=\linewidth]{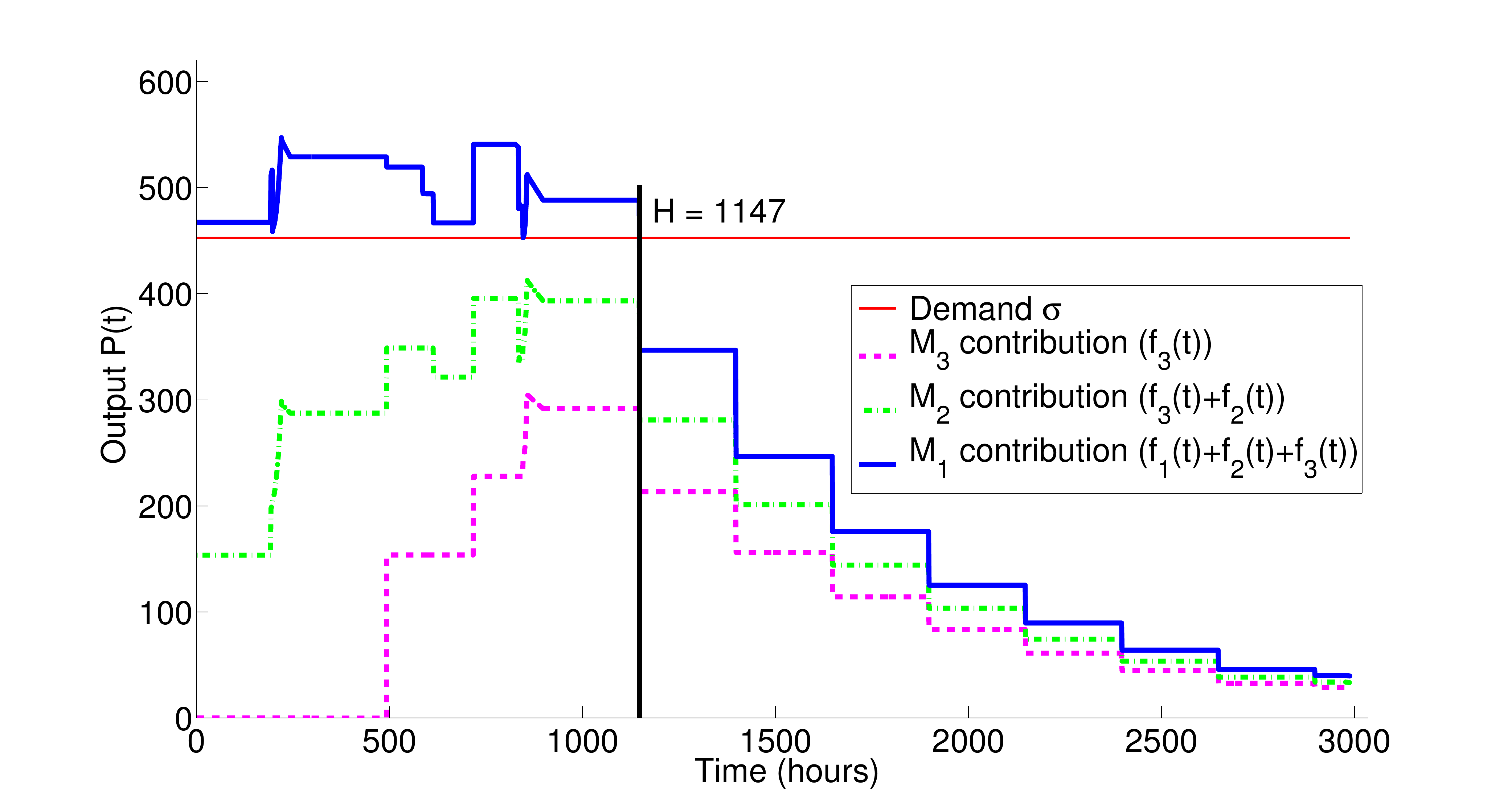}}
    \caption{Schedule\label{fig:OrdoProj}}
  \end{subfigure}
  \caption{Solution obtained with the successive projections -- $m = 3$ machines, $\sigma = 0.4\cdot\Pnom_{tot}$}
  \label{fig:SolutionProj}
\end{figure}

\begin{figure}[htb] \centering
  \begin{subfigure}[b]{\linewidth}
    \centerline{\includegraphics[scale=\scalegraphesol]{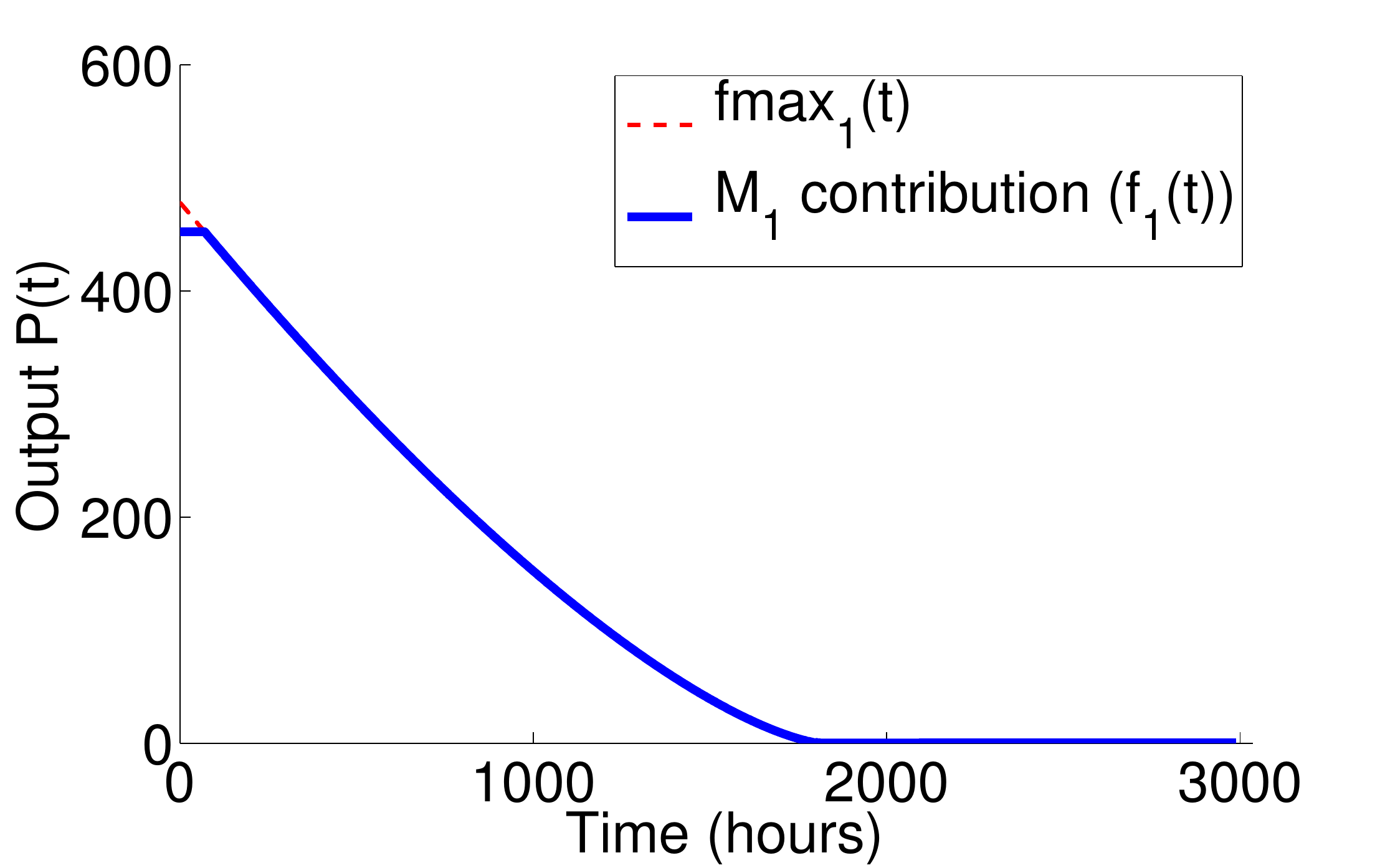}
    \includegraphics[scale=\scalegraphesol]{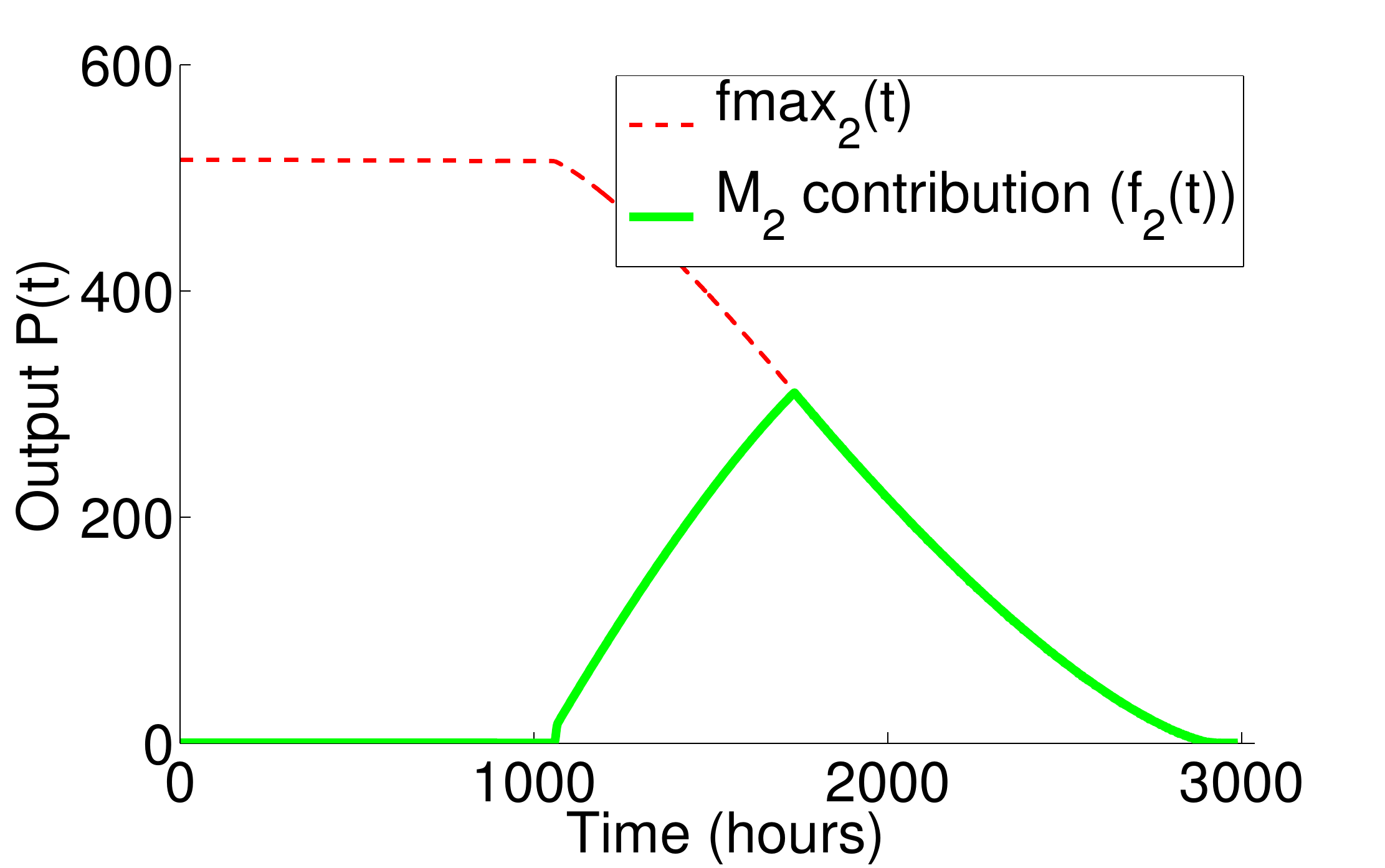}}

    \centerline{\includegraphics[scale=\scalegraphesol]{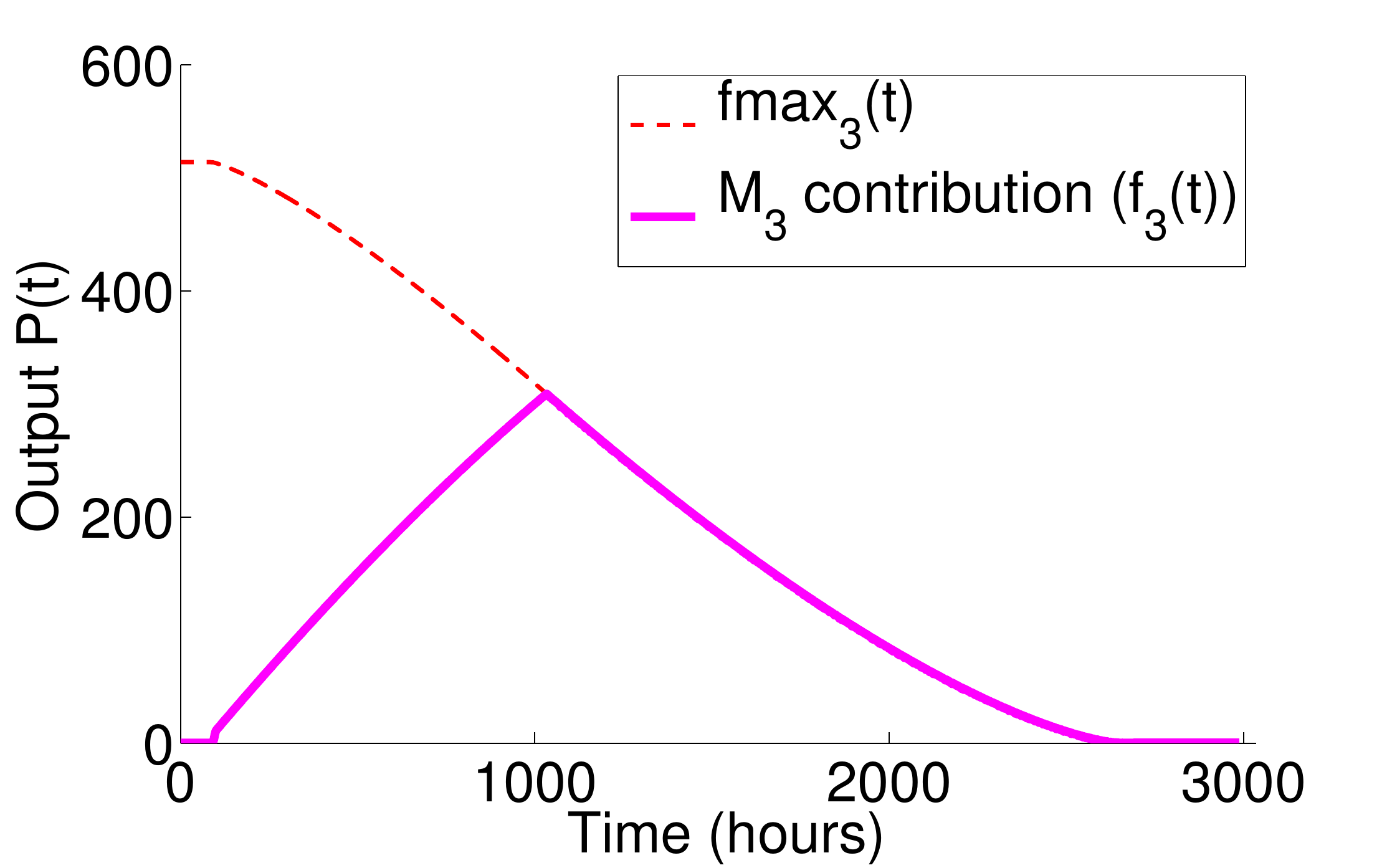}}
    \caption{Machines contributions\label{fig:detailcontributionsMPSP}}
  \end{subfigure}

  \begin{subfigure}[b]{\linewidth}
    \centerline{\includegraphics[width=\linewidth]{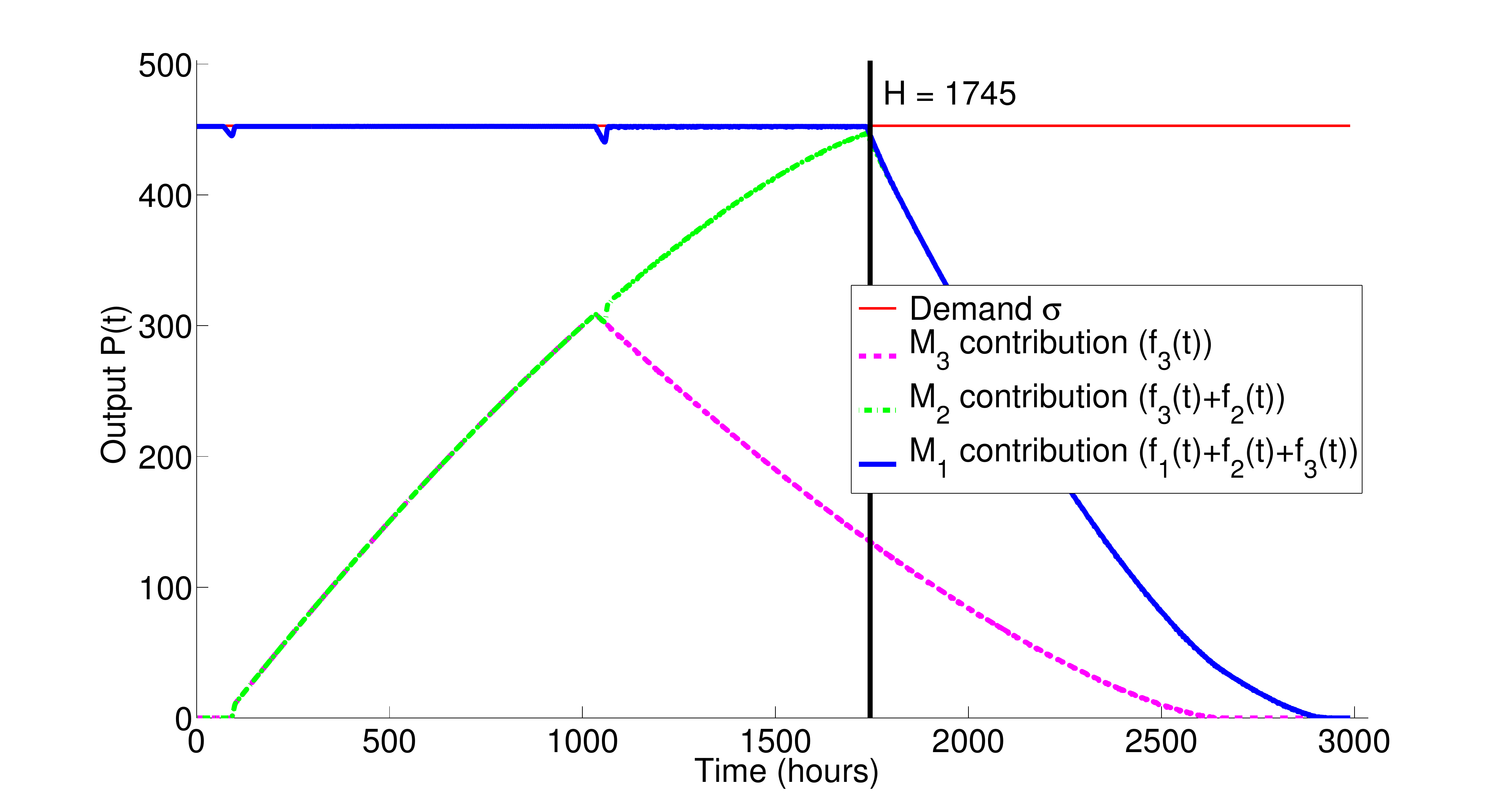}}
    \caption{Schedule\label{fig:OrdoMPSP}}
  \end{subfigure}
  \caption{Solution obtained with the Mirror-Prox-based algorithm -- $m = 3$ machines, $\sigma = 0.4\cdot\Pnom_{tot}$}
  \label{fig:SolutionMPSP}
\end{figure}

The two resolution approaches differentiate also themselves by the shape of the solutions and by their computation time. Figures~\ref{fig:SolutionProj} and~\ref{fig:SolutionMPSP} show schedules obtained respectively with the algorithm based on the successive projections and with the one based on the Mirror Prox, when considering $m = 3$ machines. Evolution of each machine contribution and of the associated maximal power output reachable, $\fmax_j$, is also shown in these figures. One can see in Figure~\ref{fig:SolutionMPSP} that the algorithm based on the Mirror Prox allows to reach a better production horizon ($H = 1745$ periods of time) than the one performing successive projections onto the sets of constraints ($H = 1147$). The Mirror Prox-based algorithm defines also a smooth use of machines, which complies with a continuous use, without sudden change of output such as those that can be seen in the solutions obtained with the successive projections (see Figure~\ref{fig:SolutionProj}).

\begin{figure}[htb]
  \centerline{\includegraphics[width=\linewidth]{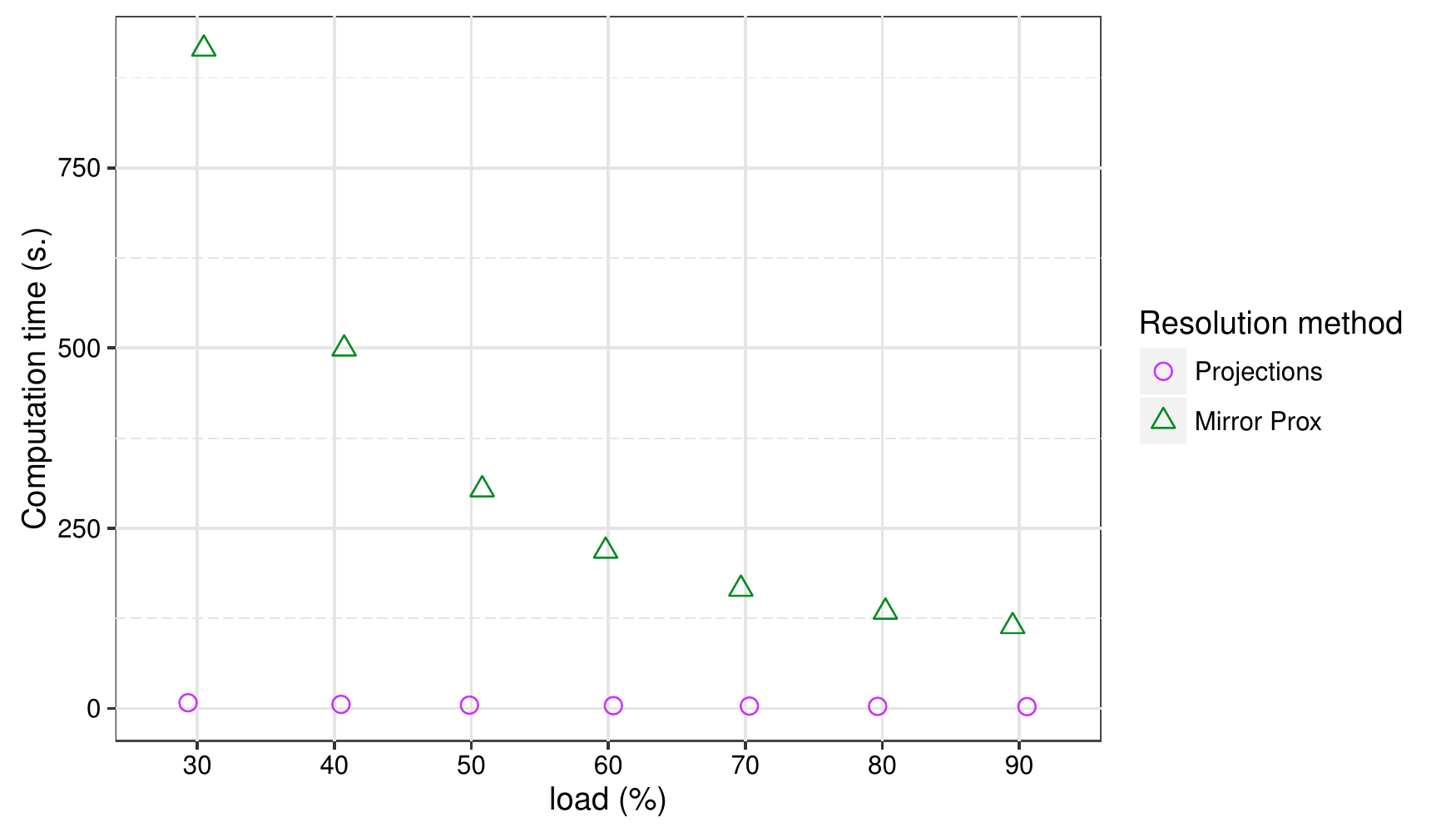}}
  \caption{Computation times -- m = 25 machines}
  \label{fig:time}
\end{figure}

The method performing successive projections onto the sets of constraints is not efficient in terms of reached production horizons, but allows to obtain solutions in very limited time (less than $9$ seconds\footnotemark[1] for all the scenarios tested when considering $25$ machines). The time needed to obtain satisfying solutions with the Mirror-Prox-based algorithm is longer ($5.6$ minutes on average), but does not exceed $17$ minutes\footnotemark[1] for $25$ machines (see Figure~\ref{fig:time}) and $35$ minutes\footnotemark[1] for $100$ machines, for all the power demands.
\footnotetext[1]{Simulations have been made using Matlab (Parameters: Processeur Intel$^{\circledR}$ Core{\tiny $^{TM}$} i5-3550 CPU@3.30GHz$\times$4, 15.6 Gio, 64 bits)}


\section{Conclusion}
\label{sec:conclusion}
A management of fuel cell systems has been proposed in a PHM framework. Decision coming within the scope of Prognostic Decision Making has been addressed considering longer time frames than those proposed so far in the literature on fuel cells. The use of convex programming has been proposed to cope with the scheduling problem of multi-stack fuel cell systems under service constraint. A mathematical formulation of the problem has been proposed as well as two different convex resolution methods performing a minimization of the objective function under constraints. First one is based on successive projections onto the sets of constraints and second one on the Mirror-Prox algorithm.

All the fuel cell properties are not observed by the solutions obtained with the proposed approaches, but this first study is promising. It shows indeed that a global resolution on the scale of the whole production horizon can be used to define the commitment of machines over time with the production horizon maximization as objective.




\textbf{Acknowledgements}:
This work has been supported by the Labex ACTION project (contract ``ANR-11-LABX-0001-01'')

\bibliographystyle{SageH}
\bibliography{CORChretien}

%
%

\end{document}